\documentclass[%
 reprint,
 superscriptaddress,
 amsmath,amssymb,
 aps,
 prl,
longbibliography
]{revtex4-1}

\usepackage{graphicx}% Include figure files
\usepackage{dcolumn}% Align table columns on decimal point
\usepackage{xcolor}
\usepackage{bm}% bold math
\usepackage[colorlinks=true,linkcolor=blue, citecolor=blue]{hyperref}%

\begin{document}

%\title{The role of oxygen atoms in the high temperature superconductors}
%from first-principles}
%\title{Inducing d-wave superconductivity through the oxygen atoms using the principles of altermagnetism in high temperature superconductors}
%\title{Applying the principles of $d$-wave altermagetism to induce $d$-wave superconductivity in the cuprates and nickelates}
%\title{Altermagnetic symmetry inducing $d$-wave superconductivity in the cuprates and nickelates}
\title{Oxygen-driven altermagnetic symmetry inducing $d$-wave superconductivity in the cuprates and nickelates}

\author{Tom G. Saunderson}
\email{thomas.saunderson@physik.uni-halle.de}
\affiliation{Institut f\"ur Physik, Martin-Luther-Universit\"at Halle-Wittenberg, D-06099 Halle (Saale), Germany} 
\author{James F. Annett}
\affiliation{H. H. Wills Physics Laboratory, University of Bristol, Tyndall Avenue, Bristol BS8-1TL, United Kingdom}
\author{Samir Lounis}%
\affiliation{Institut f\"ur Physik, Martin-Luther-Universit\"at Halle-Wittenberg, D-06099 Halle (Saale), Germany}

\date{\today}% It is always \today, today,

\begin{abstract}

Since the discovery of cuprate high-$T_c$ superconductivity, numerous theoretical frameworks have been proposed to explain its mechanism; Anderson’s RVB picture~[\href{https://doi.org/10.1126/science.235.4793.1196}{Science \textbf{235}, 1196–1198 (1987)}] and U(1) gauge theory~[\href{https://doi.org/10.1103/PhysRevLett.76.503}{Phys. Rev. Lett. \textbf{76}, 503–506 (1996)}] to name a few motivate a minimal one-band view that largely integrates out oxygen. By contrast, “altermagnetism”~[\href{https://doi.org/10.1103/PhysRevX.12.040501}{Phys. Rev. X \textbf{12}, 040501 (2022)}] yields a $d$-wave–like $k$-space magnetic texture from alternatingly rotated \emph{nonmagnetic} cages; where La$_2$CuO$_4$ (the parent of a high-$T_c$ cuprate) is a prototypical example. As a proof of principle, we show in La$_2$CuO$_4$ an alternating local pairing potential on the two Cu sublattices ($\pm s(\mathbf{r})$) produces a nodal, $d$-wave–like $\Delta(\mathbf{k})$. As orthorhombic tilts are, however, not the driver (and even suppress superconductivity in nickelates~[\href{https://doi.org/10.1038/s41586-023-06408-7}{Nature \textbf{621}, 493 (2023)})], we then show that the in-plane oxygen sublattice of CuO$_2$/NiO$_2$ layers—ubiquitous in cuprates and nickelates—intrinsically realizes the same symmetry. Imposing an \emph{oxygen-centered}, staggered $s$ pairing yields a $d$-wave gap with perfect $\mathrm{C}_4$ symmetry, demonstrated self-consistently in NdNiO$_2$ from first principles. Whilst the underlying mechanism that drives this kind of order is unclear, we provide speculations upon its origin. Further, this description of superconductivity enables the mapping of a real-space superconducting order parameter onto a lattice picture, allowing the potential for superconductivity and Hubbard physics to be treated on the same footing. 

%and providing a direct real-space mapping compatible with Hubbard models.

\end{abstract}

\maketitle

%\begin{figure*}[t!]
%\includegraphics*[width=0.9\linewidth,clip]{SupOREE_Schematic2.png}
%\caption{A schematic illustration for the superconducting orbital Rashba Edelstein effect on a non-centrosymmetric material. Similarly to the normal state, in the superconducting state there is a super-current induced orbital moment that is driven by symmetry broken orbital Rashba textures on the Fermi surface when passing a current in the $x$-direction. With this setup, the induced orbital moment will be in the y-direction. }
%\label{fig:OrbitalRashbaEdelsteinSchematic}
%\end{figure*}

\begin{figure}[h]
\includegraphics*[width=0.9\linewidth,clip]{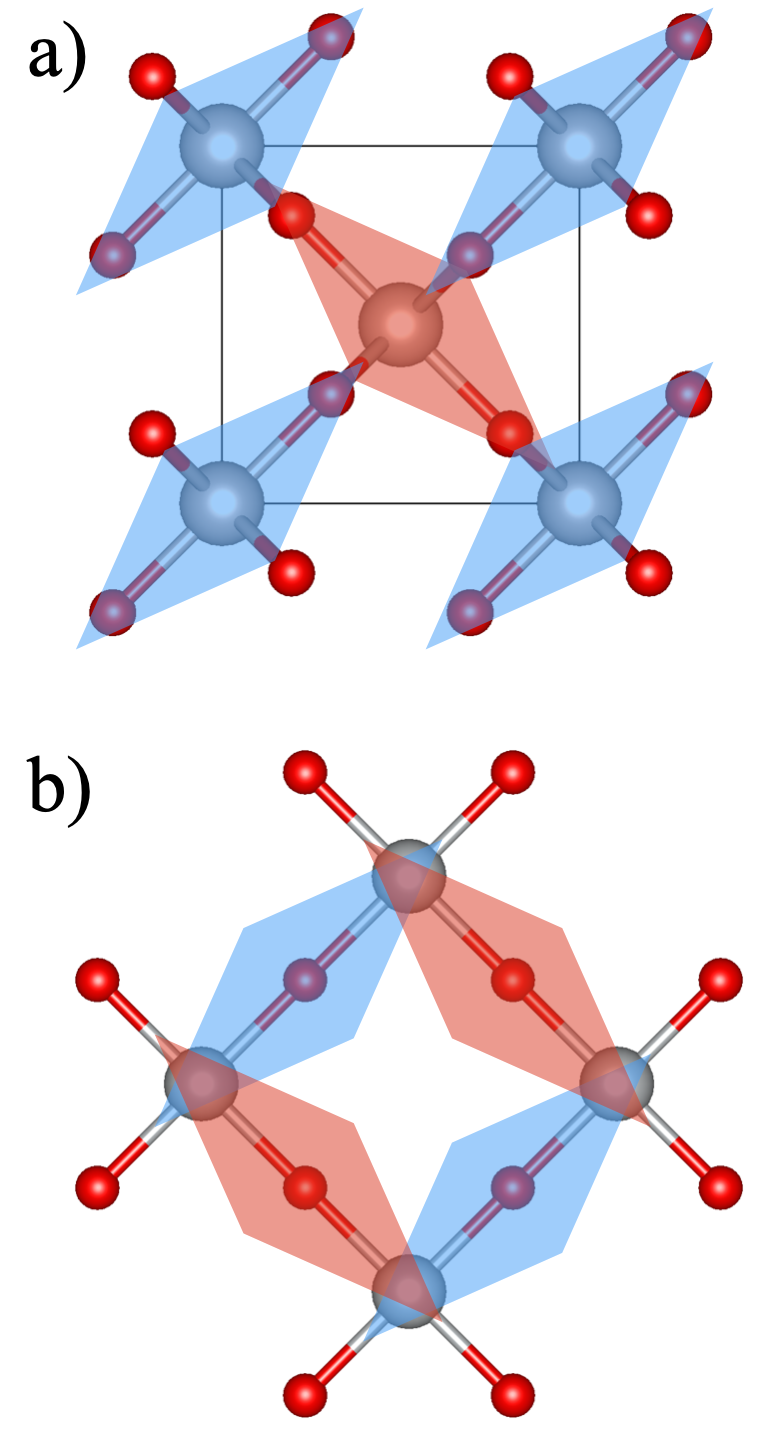}
\caption{ (a) A schematic of the crystal structure of the altermagnet RuO$_2$ highlighting the tilted octahedral cages of the crystal structure around the magnetic sites. (b) Schematic of the crystal structure of NdNiO$_2$ highlighting the NiO$_2$ plane where the oxygen sits on symmetry sites concurrent with the symmetry of the Ru atoms in RuO$_2$. }
\label{fig:AltermagnetCuprateCompare}
\end{figure}

In much the same way that the Higgs mechanism was inspired by the theory of superconductivity \cite{Anderson1963} to explain the effect of spontaneous symmetry breaking in the early universe, analogues of d-wave superconductivty were recently discovered to exist in magnetic materials, otherwise known as `altermagnets' \cite{Smejkal2022a}. Unlike ferromagnetic and antiferromagnetic materials, in order to conceptualize an altermagnet one needs to rely on the nonmagnetic atoms in the lattice, as its symmetry fundamentally arises from the structural cages of non-magnetic atoms, as displayed in Fig.~\ref{fig:AltermagnetCuprateCompare}a. The structural cages ensure that in order to map one spin onto the other one needs to perform a translation operation along with a rotation operation - whilst for antiferromagnets only the former is required. This gives rise to two different sub-lattices whose bandstructure are rotated in k-space, leading to Kramer's degeneracy being broken. It was also found that La$_2$CuO$_4$, the undoped parent of the high-$T_c$ compound La$_{2-x}$Sr$_x$CuO$_4$, exhibits an altermagnetic spin ordering due to its oxygen atoms forming tilted octahedra. These considerations motivate a fresh examination of $d$-wave superconductivity that combines altermagnetic symmetry arguments with an explicit treatment of the oxygen sublattice to sharpen our understanding of the cuprates. 

The history of theoretical modelling of the cuprates has similarly relied on focusing on the Cu atoms, most famously through single-band reductions of the CuO$_2$ planes to either a one-band Hubbard model or its strong-coupling $t\!-\!J$ descendant \cite{Anderson1987,Scalapino2012}. Although the cuprates are charge-transfer insulators in the Zaanen--Sawatzky--Allen classification \cite{Zaanen1985}, the influential mapping of the three-band $p$--$d$ (Emery) model \cite{Emery1987} onto a single effective Cu $d_{x^2-y^2}$ orbital via Zhang--Rice singlets \cite{Zhang1988} encouraged an enormous body of work that integrated out oxygen explicitly. Within this Cu-only framework---augmented at most by longer-range hoppings $t',t''$ to mimic O-mediated processes---RVB ideas, slave-particle mean-field theories, spin-fluctuation/FLEX approaches, quantum Monte Carlo, DMRG, variational Monte Carlo, and cluster extensions of DMFT/DCA all consistently identified a leading $d_{x^2-y^2}$ pairing instability and reproduced many aspects of the normal-state phenomenology \cite{Scalapino2012}. Two-band variants appeared intermittently, but in practice oxygen degrees of freedom were usually downfolded into renormalized Cu parameters \cite{Annett1990}. This Cu-centric tradition, while powerful, inevitably de-emphasized that doped holes predominantly acquire ligand-oxygen character, that superexchange is oxygen-mediated, and that several proposed symmetry-broken states (e.g., loop-current order) and materials trends (e.g., the correlation of $T_{c,\max}$ with apical-oxygen environment and effective $t'/t$) are naturally orbital---and specifically oxygen---sensitive \cite{Varma1997,Ohta1991,Pavarini2001}. More recent multi-orbital calculations that retain the O $2p$ manifold---ranging from three-band model studies \cite{Weber2011,Weber2014,Dash2019,Kowalski2021} to first-principles downfolding \cite{Kent2008,Hirayama2019,Hirayama2018} and DFT+DMFT/GW+DMFT \cite{Weber2010,Werner2015,Uchida2017} ---have begun to revisit these oxygen roles directly, setting the stage for an explicitly oxygen-centric perspective on the emergence of $d$-wave superconductivity.

%In this letter we exemplify the significance of the oxygen atoms in both copper-oxide and nickelate superconductors by employing the principles of d-wave altermagnetism to induce d-wave superconductivity. In the first instance as a proof of concept, we show that a real-space spherically symmetric pairing potential $\Delta(r)$ applied to the $Cu_1$ and $-\Delta(r)$ applied to $Cu_2$ induces a nearly $\mathrm{C}_4$ symmetric $d$-wave superconducting order parameter in the Fermi surface of La$_2$CuO$_4$. We then take a broader look at the space of high temperature superconductivity to find if there are any similarities with the nickelate family of superconductors. One key issue is that the tilted octahedra actually suppress superconductivity in the nickelates. This is overcome with the realization that the oxygen atoms sit on symmetry sites that are concurrent with altermagnetism in all cuprate and all nickelate superconductors, and in NdNiO2 the state is stable after a self-consistent calculation. 

In this Letter, we demonstrate the central role of oxygen in both cuprate and nickelate superconductors by leveraging the symmetry principles of $d$-wave altermagnetism to generate $d$-wave superconductivity. As a proof of concept, we impose a local, spherically symmetric pairing potential with opposite signs on the two oxygen-centered sublattices; $+\Delta(r)$ on Cu$_1$ and $-\Delta(r)$ on Cu$_2$—and show that this produces a nodal gap on the Fermi surface of La$_2$CuO$_4$. Whilst this gap resembles features of d-wave superconductivity, we note that it does not perfectly resemble C$_4$ symmetry. In order to search for pairing configurations that can realize C$_4$ symmetry, we broaden the scope to the nickelate family. In this family of superconductors the octahedral tilts suppress superconductivity \cite{Li2019,Sun2023}, however we observe that the oxygen sublattice in both cuprates and nickelates occupies symmetry sites that realize the altermagnetic pattern, as displayed in Fig.~\ref{fig:AltermagnetCuprateCompare}(b). Applying the oxygen-centered, alternating $s$-like pairing potential yields a perfect C$_4$ symmetric $d_{x^2-y^2}$ superconducting gap in the infinite-layer nickelate NdNiO$_2$, and the resulting superconducting order parameter remains stable under fully self-consistent calculations.

In order to understand the trends between different families of superconductors and explore their full orbital complexities, we employ a first principles Green's function-based density functional theory (DFT) method \cite{Saunderson2020,Saunderson2020b,Saunderson2022,Wu2023} which incorporates superconductivity and the full orbital degrees of freedom on the same footing. The electronic structure is obtained by solving the fully relativistic Dirac–Bogoliubov–de Gennes (DBdG) equation \cite{Csire2018}
\begin{equation}
H_{\mathrm{DBdG}}(\mathbf{r})=
\begin{pmatrix}
H_{\mathrm{D}}(\mathbf{r}) & \Delta_{\mathrm{eff}}(\mathbf{r})\\
\Delta_{\mathrm{eff}}^{*}(\mathbf{r}) & -H_{\mathrm{D}}^{*}(\mathbf{r})
\end{pmatrix},
\label{eqn:DBdG}
\end{equation}
where $H_{\mathrm{D}}(\mathbf{r})$ is the $4\times4$ Dirac Hamiltonian and $\Delta_{\mathrm{eff}}(\mathbf{r})$ is the corresponding $4\times4$ effective pairing field, taken here to be spin-singlet, however its phase can be freely varied. This first-principles superconducting formalism has been validated across a broad range of phenomena, including unconventional pairing \cite{Csire2018b,Csire2020a,Ghosh2020b}, Rashba superconductivity \cite{Rußmann2022b,Rußmann2023}, Josephson junctions \cite{Yamazaki2025}, iron-based superconductivity \cite{Reho2024}, and topological superconductivity \cite{Nyari2023,Laszloffy2023} induced by impurities \cite{Saunderson2020b,Saunderson2022,Wu2023} as well as superconductor/topological-insulator interfaces \cite{Rußmann2022a,Reho2024a,Park2020a}. Computational details of the implementation and numerical parameters are provided in the Supplementary Information~\cite{Supplementary}.

%lattice param laCuo4 \cite{Nelmes1990} 
% lattice param for NdNiO2 should be \cite{Hayward2003} but i got it from https://next-gen.materialsproject.org/materials/mp-31063?formula=NdNiO2

%We calculate the bulk orthorhombic phase (Space group Cmce) of La$_2$CuO$_4$ and the infinite layer (Space group P4/mmm) of NdNiO$_2$ in this work. 

We calculate La$_2$CuO$_4$ in its low–temperature orthorhombic phase (space group \textit{Cmce}) and NdNiO$_2$ in the infinite–layer tetragonal structure (space group \textit{P4/mmm}). In La$_2$CuO$_4$, corner–sharing CuO$_6$ octahedra undergo out–of–phase in–plane rotations (Glazer $a^{-}a^{-}c^{0}$), yielding two formula units per conventional cell and reducing the ideal $C_{4}$ symmetry of an isolated CuO$_2$ layer to $C_{2}$. This tilt pattern generates two crystallographically distinct Cu environments and splits the planar oxygen network into alternating sublattices (O$_A$/O$_B$) related by a $90^\circ$ rotation plus translation; apical oxygens complete the octahedra and modulate the Cu–O bond anisotropy. In contrast, NdNiO$_2$ adopts a simpler \textit{P4/mmm} stacking of square–planar NiO$_4$ units separated by Nd layers: apical oxygen is absent, Ni–O–Ni bonds are nearly $180^\circ$, and there is a single Ni site per primitive cell. Nevertheless, the in–plane oxygen positions form two interpenetrating sublattices related by $C_{4}$ and translation, providing a natural alternating (ligand–centered) motif. Experimental lattice parameters and internal coordinates are taken from Refs.~\cite{Nelmes1990,Hayward2003}.

\begin{figure}[h]
\includegraphics*[width=0.95\linewidth,clip]{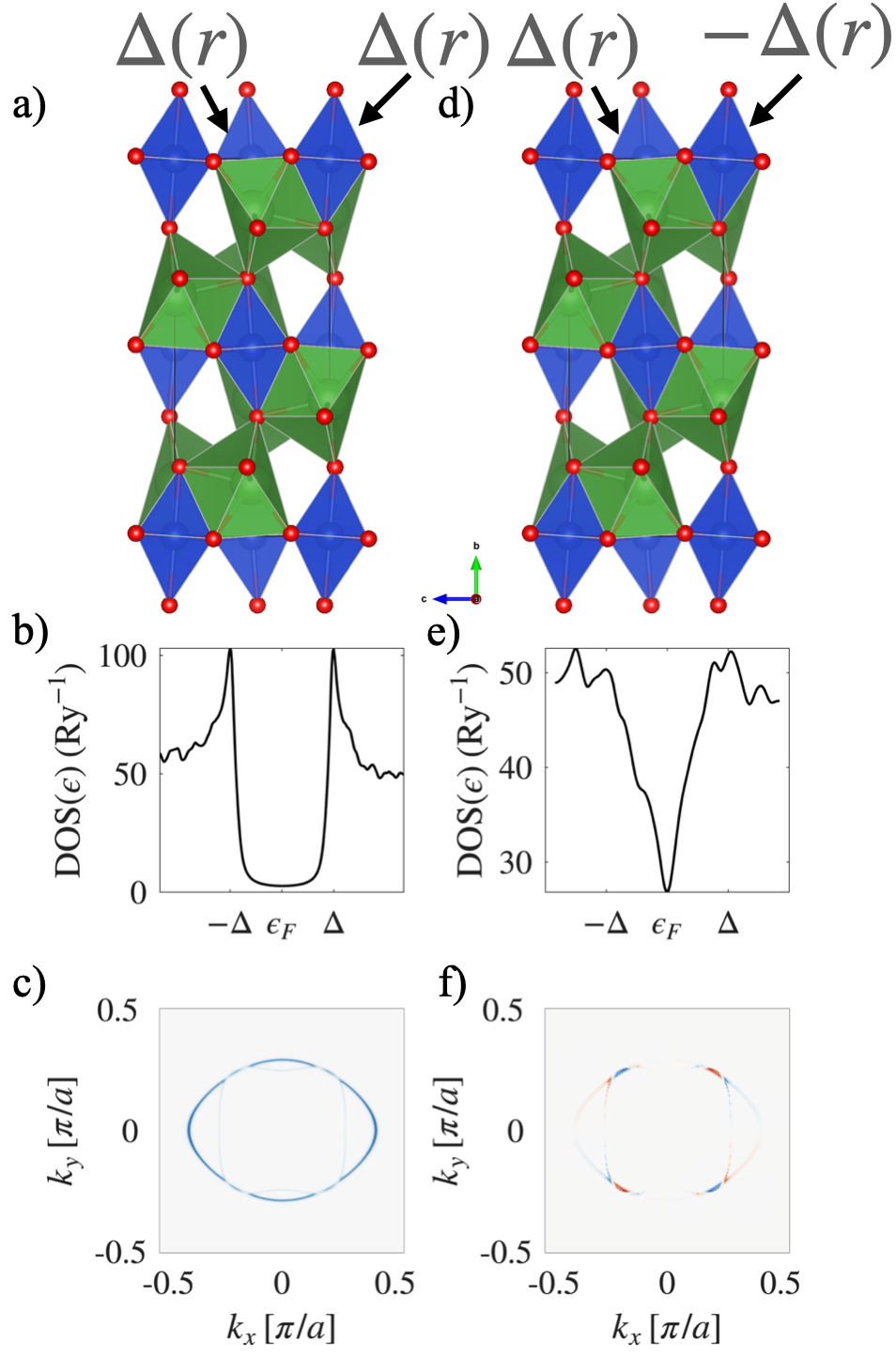}
\caption{ (a, d) A schematic of the crystal structure of La$_2$CuO$_4$ highlighting the tilted octahedral cages of the crystal structure, with the highlighted pairing potential applied to the Cu atom sites within. (b, e) Density of states around the Fermi level depicting the superconducting gap in both scenarios of choice of pairing parameter. (c, f) Spectral function of the anomalous density around the coherence peak of (b) and (e). It is clear that in (c) the plot is isotropic as one rotates around the bandstructure, whilst in (f) there are clear 4 nodal points. }
\label{fig:La2CuO4_plot}
\end{figure}

We begin with La$_2$CuO$_4$. As a baseline, we impose a uniform on-site pairing field of equal magnitude and phase on the two Cu sites in the unit cell (residing in distinct octahedra), as labeled in Fig.~\ref{fig:La2CuO4_plot}(a). The resulting density of states (DOS) displays a pronounced U-shaped gap, Fig.~\ref{fig:La2CuO4_plot}(b), consistent with a nodeless, $s$-like response. Although the corresponding anomalous density on the Fermi surface, Fig.~\ref{fig:La2CuO4_plot}(c), exhibits nontrivial structure, it does not reveal a clear, symmetry-enforced $d$-wave sign pattern.

We then introduce a staggered pairing configuration, applying $+\Delta(r)$ on Cu$_1$ and $-\Delta(r)$ on Cu$_2$, as depicted in Fig.~\ref{fig:La2CuO4_plot}(d). This immediately produces a V-shaped DOS, Fig.~\ref{fig:La2CuO4_plot}(e), indicative of a nodal gap. In momentum space, the anomalous density now shows four nodal directions characteristic of a $d$-wave order parameter [Fig.~\ref{fig:La2CuO4_plot}(f)]. The apparent reduction to a $C_2$-like pattern arises from the lowered crystallographic symmetry (e.g., octahedral tilts) rather than from the pairing symmetry itself; in an idealized tetragonal reference, the structure approaches $C_4$. It is also worth noting that previous experiments \cite{Radaelli1994,Naito2018} have determined that whilst the orthorhombic structure does alter the symmetry, the tilting does not have a significant effect on the superconductivity. 

%Now we advance to NdNiO$_2$. This material is similar to La2CuO4 in the sense that the tilting is unrelated to the superconductivity. One main difference is that in NdNiO$_2$, the tilting must be removed in order that the superconducting order emerges. In recent literature it appears that  This material has a significantly more basic crystal structure compared to the full complexity of La$_2$CuO$_4$. On first inspection, it is clear that the Ni site does not have the same rotated octahedral cages, in fact the structure is too simple for any possible complicated symmetry for the Ni site Fig.~3(a). Alternatively, inspecting the oxygen site, there exists a zig-zag structure coming from the Ni atoms surrounding the oxygen sites which alternate from one oxygen site to the next, giving rise to the classic altermagnetic spin symmetry site of a conventional d-wave altermagnet. It has not, however, been noticed before as the oxygen atom would never be magnetic. As we can see in Fig.~3(b) the same V-shaped gap is present in the density of states, and the corresponing d-wave order parameter on the Fermi surface in Fig.~3(c) with perfect $C_4$ symmetry is returned. This calculation is performed self-consistently. 

We now turn to the infinite-layer nickelate NdNiO$_2$. In contrast to La$_2$CuO$_4$, where octahedral tilts mainly complicate the crystallography without being essential to superconductivity, the nickelates exhibit a stronger sensitivity to lattice distortions: perovskite-like tilt patterns tend to suppress pairing, and the emergence of superconductivity correlates with the square–planar, untilted (infinite-layer) structure. Consistent with this, we remove octahedral tilts to restore the high-symmetry reference for NdNiO$_2$ prior to introducing pairing.

\begin{figure}[h]
\includegraphics[width=0.65\linewidth,clip]{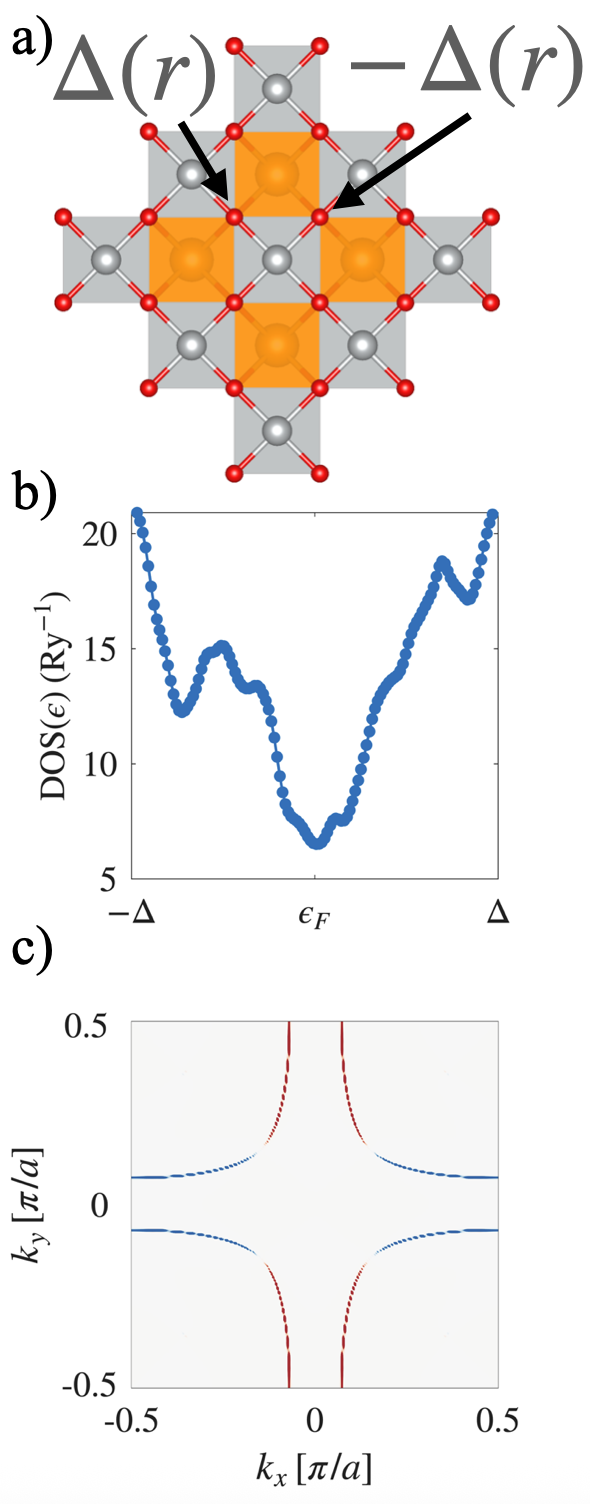}
\caption{ (a) A schematic of the crystal structure of La$_2$CuO$_4$ highlighting the tilted octahedral cages of the crystal structure, with the highlighted pairing potential applied to the Cu atom sites within. (b) Density of states around the Fermi level depicting the superconducting gap in both scenarios of choice of pairing parameter. (c) Spectral function of the anomalous density around the coherence peak of (b). It is clear that in (c) the $C_4$ symmetry is achieved. }
\label{fig:NdNiO2_plot}
\end{figure}

Structurally, the Ni site in NdNiO$_2$ is too simple to host the rotated-octahedra motifs that underpin $k$-space rotations in altermagnets [Fig.~\ref{fig:NdNiO2_plot}(a)]. By contrast, the in-plane oxygen sublattice has two surrounding Ni atoms whose positions are rotated by 45 degrees from one O site to the next, naturally realizing the symmetry locus of a conventional $d$-wave altermagnet. Although oxygen is typically treated as nonmagnetic, the same symmetry can be harnessed by an oxygen-centered, alternating $s$-like pairing field. Implementing this oxygen-staggered pairing yields a V-shaped superconducting gap in the DOS [Fig.~\ref{fig:NdNiO2_plot}(b)] and a $d$-wave order parameter on the Fermi surface with perfect $C_4$ symmetry [Fig.~\ref{fig:NdNiO2_plot}(c)]. All results reported here are obtained from fully self-consistent calculations.

One possible explaination that could manifest the alternating, oxygen-centered pairing field employed above can be rationalized without committing to a purely “phononless” mechanism. In the linearized gap equation,
\begin{equation}
\Delta(\mathbf{k}) = - \sum_{\mathbf{k}'} V(\mathbf{k}-\mathbf{k}') \,\chi(\mathbf{k}',T)\, \Delta(\mathbf{k}'),
\label{eq:gap_equation}
\end{equation}
the net pairing kernel decomposes as $V(\mathbf{q}) \!=\! V_{\mathrm{ph}}(\mathbf{q}) - \mu^{*}$, where the Coulomb pseudopotential $\mu^{*}$ is predominantly local (hence only weakly $\mathbf{q}$-dependent). Consequently, $\mu^{*}$ penalizes the $s$-wave channel most strongly, while it couples only weakly to sign-changing form factors whose Fermi-surface average vanishes, such as $d_{x^{2}-y^{2}}$. If, in addition, the electron–phonon interaction carried by oxygen modes is \emph{momentum selective}—for example, enhanced for bond-stretching/buckling/breathing vibrations that produce a strongly $\mathbf{q}$-dependent $V_{\mathrm{ph}}(\mathbf{q})$—then the leading eigenfunction of Eq.~(\ref{eq:gap_equation}) can naturally acquire sign changes, favoring $s_{\pm}$ or $d$-wave states over a conventional $s$-wave even when $V_{\mathrm{ph}}$ is attractive in multiple channels. In real space, such a momentum-structured attraction projects onto the oxygen sublattice as an alternating nearest-neighbor pairing amplitude: adjacent O sites experience opposite phases because their Cu/Ni coordination alternates, precisely the altermagnetic motif identified above. This mapping explains why an oxygen-centered, staggered $s$-like $\Delta_{\mathrm{eff}}(\mathbf{r})$ generates a $d$-wave gap $\Delta(\mathbf{k})$ with (nearly) restored $C_{4}$ symmetry on the Fermi surface, while remaining comparatively insensitive to the local repulsion $\mu^{*}$. We emphasize that this is a plausible route rather than a worked-out microscopic theory; a quantitative treatment of the oxygen-mode spectrum, its coupling matrix elements, and the competition between $s$, $s_{\pm}$, and $d$ channels is left for future work.

To conclude, we have shown that La$_2$CuO$_4$ can host a nodal gap with similarities to $d$-wave superconductivity generated from the Cu sites by enforcing an alternating pairing potential between the two Cu-centered octahedra. However, this Cu-based route is not ubiquitous and depends on the tilting octahedral cages of the La2CuO4, which have been previously experimentally proven to not be the fundamental driver of the superconducting order parameter in the cuprates. By contrast, in NdNiO$_2$ we identify a previously unrecognized, \emph{oxygen-sublattice} altermagnetic symmetry: the alternating Ni coordination of in-plane O sites naturally encodes the sign structure required for a $d$-wave state. While our microscopic rationale for why oxygen modes select this channel remains speculative, the corresponding oxygen-centered, staggered pairing field yields a perfect C$_4$ symmetric $d_{x^2-y^2}$ gap that is \emph{self-consistently} stable within our DBdG framework.

Because the same oxygen-site symmetry motif is intrinsic to \emph{all} CuO$_2$ and NiO$_2$ planes, this provides a general and compelling route to $d$-wave superconductivity that does not rely on material-specific fine tuning at the transition-metal site. Crucially, the construction is formulated in real space, allowing a direct mapping onto Hubbard-style lattice models as an alternating ligand-centered pairing amplitude, rather than appending a $\mathbf{k}$-space gap in a second-variational step.

\begin{acknowledgments}
The authors appreciate fruitful discussions with Prof. Yuriy Mokrousov and Dr. Martin Gradhand. 
%This work was supported by '...'  
\end{acknowledgments}

\bibliography{CuprateNikelatePaper}

\begin{thebibliography}{45}%
\makeatletter
\providecommand \@ifxundefined [1]{%
 \@ifx{#1\undefined}
}%
\providecommand \@ifnum [1]{%
 \ifnum #1\expandafter \@firstoftwo
 \else \expandafter \@secondoftwo
 \fi
}%
\providecommand \@ifx [1]{%
 \ifx #1\expandafter \@firstoftwo
 \else \expandafter \@secondoftwo
 \fi
}%
\providecommand \natexlab [1]{#1}%
\providecommand \enquote  [1]{``#1''}%
\providecommand \bibnamefont  [1]{#1}%
\providecommand \bibfnamefont [1]{#1}%
\providecommand \citenamefont [1]{#1}%
\providecommand \href@noop [0]{\@secondoftwo}%
\providecommand \href [0]{\begingroup \@sanitize@url \@href}%
\providecommand \@href[1]{\@@startlink{#1}\@@href}%
\providecommand \@@href[1]{\endgroup#1\@@endlink}%
\providecommand \@sanitize@url [0]{\catcode `\\12\catcode `\$12\catcode `\&12\catcode `\#12\catcode `\^12\catcode `\_12\catcode `\%12\relax}%
\providecommand \@@startlink[1]{}%
\providecommand \@@endlink[0]{}%
\providecommand \url  [0]{\begingroup\@sanitize@url \@url }%
\providecommand \@url [1]{\endgroup\@href {#1}{\urlprefix }}%
\providecommand \urlprefix  [0]{URL }%
\providecommand \Eprint [0]{\href }%
\providecommand \doibase [0]{http://dx.doi.org/}%
\providecommand \selectlanguage [0]{\@gobble}%
\providecommand \bibinfo  [0]{\@secondoftwo}%
\providecommand \bibfield  [0]{\@secondoftwo}%
\providecommand \translation [1]{[#1]}%
\providecommand \BibitemOpen [0]{}%
\providecommand \bibitemStop [0]{}%
\providecommand \bibitemNoStop [0]{.\EOS\space}%
\providecommand \EOS [0]{\spacefactor3000\relax}%
\providecommand \BibitemShut  [1]{\csname bibitem#1\endcsname}%
\let\auto@bib@innerbib\@empty
%</preamble>
\bibitem [{\citenamefont {Anderson}(1963)}]{Anderson1963}%
  \BibitemOpen
  \bibfield  {author} {\bibinfo {author} {\bibfnamefont {P.~W.}\ \bibnamefont {Anderson}},\ }\bibfield  {title} {\enquote {\bibinfo {title} {{Plasmons, gauge invariance, and mass}},}\ }\href {\doibase 10.1103/PhysRev.130.439} {\bibfield  {journal} {\bibinfo  {journal} {Phys. Rev.}\ }\textbf {\bibinfo {volume} {130}},\ \bibinfo {pages} {439--442} (\bibinfo {year} {1963})}\BibitemShut {NoStop}%
\bibitem [{\citenamefont {{\v{S}}mejkal}\ \emph {et~al.}(2022)\citenamefont {{\v{S}}mejkal}, \citenamefont {Sinova},\ and\ \citenamefont {Jungwirth}}]{Smejkal2022a}%
  \BibitemOpen
  \bibfield  {author} {\bibinfo {author} {\bibfnamefont {Libor}\ \bibnamefont {{\v{S}}mejkal}}, \bibinfo {author} {\bibfnamefont {Jairo}\ \bibnamefont {Sinova}}, \ and\ \bibinfo {author} {\bibfnamefont {Tomas}\ \bibnamefont {Jungwirth}},\ }\bibfield  {title} {\enquote {\bibinfo {title} {{Emerging Research Landscape of Altermagnetism}},}\ }\href {\doibase 10.1103/PHYSREVX.12.040501/FIGURES/14/MEDIUM} {\bibfield  {journal} {\bibinfo  {journal} {Phys. Rev. X}\ }\textbf {\bibinfo {volume} {12}},\ \bibinfo {pages} {040501} (\bibinfo {year} {2022})},\ \Eprint {http://arxiv.org/abs/2204.10844} {arXiv:2204.10844} \BibitemShut {NoStop}%
\bibitem [{\citenamefont {Anderson}(1987)}]{Anderson1987}%
  \BibitemOpen
  \bibfield  {author} {\bibinfo {author} {\bibfnamefont {P.~W.}\ \bibnamefont {Anderson}},\ }\bibfield  {title} {\enquote {\bibinfo {title} {{The Resonating Valence Bond State in La$_2$CuO$_4$ and Superconductivity}},}\ }\href@noop {} {\bibfield  {journal} {\bibinfo  {journal} {Science (80-. ).}\ }\textbf {\bibinfo {volume} {235}},\ \bibinfo {pages} {1196--1198} (\bibinfo {year} {1987})}\BibitemShut {NoStop}%
\bibitem [{\citenamefont {Scalapino}(2012)}]{Scalapino2012}%
  \BibitemOpen
  \bibfield  {author} {\bibinfo {author} {\bibfnamefont {D.~J.}\ \bibnamefont {Scalapino}},\ }\bibfield  {title} {\enquote {\bibinfo {title} {{A common thread: The pairing interaction for unconventional superconductors}},}\ }\href {\doibase 10.1103/RevModPhys.84.1383} {\bibfield  {journal} {\bibinfo  {journal} {Rev. Mod. Phys.}\ }\textbf {\bibinfo {volume} {84}},\ \bibinfo {pages} {1383--1417} (\bibinfo {year} {2012})},\ \Eprint {http://arxiv.org/abs/1207.4093} {arXiv:1207.4093} \BibitemShut {NoStop}%
\bibitem [{\citenamefont {Zaanen}\ \emph {et~al.}(1985)\citenamefont {Zaanen}, \citenamefont {Sawatzky},\ and\ \citenamefont {Allen}}]{Zaanen1985}%
  \BibitemOpen
  \bibfield  {author} {\bibinfo {author} {\bibfnamefont {J.}~\bibnamefont {Zaanen}}, \bibinfo {author} {\bibfnamefont {G.~A.}\ \bibnamefont {Sawatzky}}, \ and\ \bibinfo {author} {\bibfnamefont {J.~W.}\ \bibnamefont {Allen}},\ }\bibfield  {title} {\enquote {\bibinfo {title} {{Band gaps and electronic structure of transition-metal compounds}},}\ }\href {\doibase 10.1103/PhysRevLett.55.418} {\bibfield  {journal} {\bibinfo  {journal} {Phys. Rev. Lett.}\ }\textbf {\bibinfo {volume} {55}},\ \bibinfo {pages} {418} (\bibinfo {year} {1985})}\BibitemShut {NoStop}%
\bibitem [{\citenamefont {Emery}(1987)}]{Emery1987}%
  \BibitemOpen
  \bibfield  {author} {\bibinfo {author} {\bibfnamefont {V.~J.}\ \bibnamefont {Emery}},\ }\bibfield  {title} {\enquote {\bibinfo {title} {{Theory of High-Tc superconductivity in Oxides}},}\ }\href {\doibase 10.1103/PhysRevLett.58.2794} {\bibfield  {journal} {\bibinfo  {journal} {Phys. Rev. Lett.}\ }\textbf {\bibinfo {volume} {58}},\ \bibinfo {pages} {2794} (\bibinfo {year} {1987})}\BibitemShut {NoStop}%
\bibitem [{\citenamefont {Zhang}\ and\ \citenamefont {Rice}(1988)}]{Zhang1988}%
  \BibitemOpen
  \bibfield  {author} {\bibinfo {author} {\bibfnamefont {F.~C.}\ \bibnamefont {Zhang}}\ and\ \bibinfo {author} {\bibfnamefont {T.~M.}\ \bibnamefont {Rice}},\ }\bibfield  {title} {\enquote {\bibinfo {title} {{Effective Hamiltonian for the superconducting Cu oxides}},}\ }\href {\doibase 10.1103/PhysRevB.37.3759} {\bibfield  {journal} {\bibinfo  {journal} {Phys. Rev. B}\ }\textbf {\bibinfo {volume} {37}},\ \bibinfo {pages} {3759} (\bibinfo {year} {1988})}\BibitemShut {NoStop}%
\bibitem [{\citenamefont {Annett}\ and\ \citenamefont {Martin}(1990)}]{Annett1990}%
  \BibitemOpen
  \bibfield  {author} {\bibinfo {author} {\bibfnamefont {James~F.}\ \bibnamefont {Annett}}\ and\ \bibinfo {author} {\bibfnamefont {Richard~M.}\ \bibnamefont {Martin}},\ }\bibfield  {title} {\enquote {\bibinfo {title} {{Two-band Hamiltonian for CuO2 planes}},}\ }\href {\doibase 10.1103/PhysRevB.42.3929} {\bibfield  {journal} {\bibinfo  {journal} {Phys. Rev. B}\ }\textbf {\bibinfo {volume} {42}},\ \bibinfo {pages} {3929--3934} (\bibinfo {year} {1990})}\BibitemShut {NoStop}%
\bibitem [{\citenamefont {Varma}(1997)}]{Varma1997}%
  \BibitemOpen
  \bibfield  {author} {\bibinfo {author} {\bibfnamefont {C.}~\bibnamefont {Varma}},\ }\bibfield  {title} {\enquote {\bibinfo {title} {{Non-Fermi-liquid states and pairing instability of a general model of copper oxide metals}},}\ }\href {\doibase 10.1103/PhysRevB.55.14554} {\bibfield  {journal} {\bibinfo  {journal} {Phys. Rev. B}\ }\textbf {\bibinfo {volume} {55}},\ \bibinfo {pages} {14554} (\bibinfo {year} {1997})}\BibitemShut {NoStop}%
\bibitem [{\citenamefont {Ohta}\ \emph {et~al.}(1991)\citenamefont {Ohta}, \citenamefont {Tohyama},\ and\ \citenamefont {Maekawa}}]{Ohta1991}%
  \BibitemOpen
  \bibfield  {author} {\bibinfo {author} {\bibfnamefont {Y.}~\bibnamefont {Ohta}}, \bibinfo {author} {\bibfnamefont {T.}~\bibnamefont {Tohyama}}, \ and\ \bibinfo {author} {\bibfnamefont {S.}~\bibnamefont {Maekawa}},\ }\bibfield  {title} {\enquote {\bibinfo {title} {{Apex oxygen and critical temperature in copper oxide superconductors: Universal correlation with the stability of local singlets}},}\ }\href {\doibase 10.1103/PhysRevB.43.2968} {\bibfield  {journal} {\bibinfo  {journal} {Phys. Rev. B}\ }\textbf {\bibinfo {volume} {43}},\ \bibinfo {pages} {2968} (\bibinfo {year} {1991})}\BibitemShut {NoStop}%
\bibitem [{\citenamefont {Pavarini}\ \emph {et~al.}(2001)\citenamefont {Pavarini}, \citenamefont {Dasgupta}, \citenamefont {Saha-Dasgupta}, \citenamefont {Jepsen},\ and\ \citenamefont {Andersen}}]{Pavarini2001}%
  \BibitemOpen
  \bibfield  {author} {\bibinfo {author} {\bibfnamefont {E.}~\bibnamefont {Pavarini}}, \bibinfo {author} {\bibfnamefont {I.}~\bibnamefont {Dasgupta}}, \bibinfo {author} {\bibfnamefont {T.}~\bibnamefont {Saha-Dasgupta}}, \bibinfo {author} {\bibfnamefont {O.}~\bibnamefont {Jepsen}}, \ and\ \bibinfo {author} {\bibfnamefont {O.~K.}\ \bibnamefont {Andersen}},\ }\bibfield  {title} {\enquote {\bibinfo {title} {{Band-Structure Trend in Hole-Doped Cuprates and Correlation with Tc}},}\ }\href {\doibase 10.1103/PhysRevLett.87.047003} {\bibfield  {journal} {\bibinfo  {journal} {Phys. Rev. Lett.}\ }\textbf {\bibinfo {volume} {87}},\ \bibinfo {pages} {047003} (\bibinfo {year} {2001})}\BibitemShut {NoStop}%
\bibitem [{\citenamefont {Weber}\ \emph {et~al.}(2011)\citenamefont {Weber}, \citenamefont {Yee}, \citenamefont {Haule},\ and\ \citenamefont {Kotliar}}]{Weber2011}%
  \BibitemOpen
  \bibfield  {author} {\bibinfo {author} {\bibfnamefont {C.}~\bibnamefont {Weber}}, \bibinfo {author} {\bibfnamefont {C.}~\bibnamefont {Yee}}, \bibinfo {author} {\bibfnamefont {K.}~\bibnamefont {Haule}}, \ and\ \bibinfo {author} {\bibfnamefont {G.}~\bibnamefont {Kotliar}},\ }\bibfield  {title} {\enquote {\bibinfo {title} {{Scaling of the transition temperature of hole-doped cuprate superconductors with the charge-transfer energy}},}\ }\href {\doibase 10.1209/0295-5075/100/37001} {\bibfield  {journal} {\bibinfo  {journal} {EPL}\ }\textbf {\bibinfo {volume} {100}},\ \bibinfo {pages} {37001} (\bibinfo {year} {2011})}\BibitemShut {NoStop}%
\bibitem [{\citenamefont {Weber}\ \emph {et~al.}(2014)\citenamefont {Weber}, \citenamefont {Giamarchi},\ and\ \citenamefont {Varma}}]{Weber2014}%
  \BibitemOpen
  \bibfield  {author} {\bibinfo {author} {\bibfnamefont {C{\'{e}}dric}\ \bibnamefont {Weber}}, \bibinfo {author} {\bibfnamefont {T.}~\bibnamefont {Giamarchi}}, \ and\ \bibinfo {author} {\bibfnamefont {C.~M.}\ \bibnamefont {Varma}},\ }\bibfield  {title} {\enquote {\bibinfo {title} {{Phase diagram of a three-orbital model for high- T c cuprate superconductors}},}\ }\href {\doibase 10.1103/PHYSREVLETT.112.117001/FIGURES/3/MEDIUM} {\bibfield  {journal} {\bibinfo  {journal} {Phys. Rev. Lett.}\ }\textbf {\bibinfo {volume} {112}},\ \bibinfo {pages} {117001} (\bibinfo {year} {2014})}\BibitemShut {NoStop}%
\bibitem [{\citenamefont {Dash}\ and\ \citenamefont {S{\'{e}}n{\'{e}}chal}(2019)}]{Dash2019}%
  \BibitemOpen
  \bibfield  {author} {\bibinfo {author} {\bibfnamefont {S.~S.}\ \bibnamefont {Dash}}\ and\ \bibinfo {author} {\bibfnamefont {D.}~\bibnamefont {S{\'{e}}n{\'{e}}chal}},\ }\bibfield  {title} {\enquote {\bibinfo {title} {{Pseudogap transition within the superconducting phase in the three-band Hubbard model}},}\ }\href {\doibase 10.1103/PHYSREVB.100.214509/FIGURES/9/THUMBNAIL} {\bibfield  {journal} {\bibinfo  {journal} {Phys. Rev. B}\ }\textbf {\bibinfo {volume} {100}},\ \bibinfo {pages} {214509} (\bibinfo {year} {2019})},\ \Eprint {http://arxiv.org/abs/1910.03522} {arXiv:1910.03522} \BibitemShut {NoStop}%
\bibitem [{\citenamefont {Kowalski}\ \emph {et~al.}(2021)\citenamefont {Kowalski}, \citenamefont {Dash}, \citenamefont {S{\'{e}}mon}, \citenamefont {S{\'{e}}n{\'{e}}chal},\ and\ \citenamefont {Tremblay}}]{Kowalski2021}%
  \BibitemOpen
  \bibfield  {author} {\bibinfo {author} {\bibfnamefont {Nicolas}\ \bibnamefont {Kowalski}}, \bibinfo {author} {\bibfnamefont {Sidhartha~Shankar}\ \bibnamefont {Dash}}, \bibinfo {author} {\bibfnamefont {Patrick}\ \bibnamefont {S{\'{e}}mon}}, \bibinfo {author} {\bibfnamefont {David}\ \bibnamefont {S{\'{e}}n{\'{e}}chal}}, \ and\ \bibinfo {author} {\bibfnamefont {Andr{\'{e}}~Marie}\ \bibnamefont {Tremblay}},\ }\bibfield  {title} {\enquote {\bibinfo {title} {{Oxygen hole content, charge-transfer gap, covalency, and cuprate superconductivity}},}\ }\href {\doibase 10.1073/PNAS.2106476118/SUPPL_FILE/PNAS.2106476118.SAPP.PDF} {\bibfield  {journal} {\bibinfo  {journal} {PNAS}\ }\textbf {\bibinfo {volume} {118}},\ \bibinfo {pages} {e2106476118} (\bibinfo {year} {2021})},\ \Eprint {http://arxiv.org/abs/2104.07087} {arXiv:2104.07087} \BibitemShut {NoStop}%
\bibitem [{\citenamefont {Kent}\ \emph {et~al.}(2008)\citenamefont {Kent}, \citenamefont {Saha-Dasgupta}, \citenamefont {Jepsen}, \citenamefont {Andersen}, \citenamefont {MacRidin}, \citenamefont {Maier}, \citenamefont {Jarrell},\ and\ \citenamefont {Schulthess}}]{Kent2008}%
  \BibitemOpen
  \bibfield  {author} {\bibinfo {author} {\bibfnamefont {P.~R.C.}\ \bibnamefont {Kent}}, \bibinfo {author} {\bibfnamefont {T.}~\bibnamefont {Saha-Dasgupta}}, \bibinfo {author} {\bibfnamefont {O.}~\bibnamefont {Jepsen}}, \bibinfo {author} {\bibfnamefont {O.~K.}\ \bibnamefont {Andersen}}, \bibinfo {author} {\bibfnamefont {A.}~\bibnamefont {MacRidin}}, \bibinfo {author} {\bibfnamefont {T.~A.}\ \bibnamefont {Maier}}, \bibinfo {author} {\bibfnamefont {M.}~\bibnamefont {Jarrell}}, \ and\ \bibinfo {author} {\bibfnamefont {T.~C.}\ \bibnamefont {Schulthess}},\ }\bibfield  {title} {\enquote {\bibinfo {title} {{Combined density functional and dynamical cluster quantum Monte Carlo calculations of the three-band Hubbard model for hole-doped cuprate superconductors}},}\ }\href {\doibase 10.1103/PHYSREVB.78.035132/FIGURES/7/THUMBNAIL} {\bibfield  {journal} {\bibinfo  {journal} {Phys. Rev. B - Condens. Matter Mater. Phys.}\ }\textbf {\bibinfo {volume} {78}},\ \bibinfo {pages} {035132} (\bibinfo {year} {2008})},\ \Eprint {http://arxiv.org/abs/0806.3770} {arXiv:0806.3770} \BibitemShut {NoStop}%
\bibitem [{\citenamefont {Hirayama}\ \emph {et~al.}(2019)\citenamefont {Hirayama}, \citenamefont {Misawa}, \citenamefont {Ohgoe}, \citenamefont {Yamaji},\ and\ \citenamefont {Imada}}]{Hirayama2019}%
  \BibitemOpen
  \bibfield  {author} {\bibinfo {author} {\bibfnamefont {Motoaki}\ \bibnamefont {Hirayama}}, \bibinfo {author} {\bibfnamefont {Takahiro}\ \bibnamefont {Misawa}}, \bibinfo {author} {\bibfnamefont {Takahiro}\ \bibnamefont {Ohgoe}}, \bibinfo {author} {\bibfnamefont {Youhei}\ \bibnamefont {Yamaji}}, \ and\ \bibinfo {author} {\bibfnamefont {Masatoshi}\ \bibnamefont {Imada}},\ }\bibfield  {title} {\enquote {\bibinfo {title} {{Effective Hamiltonian for cuprate superconductors derived from multiscale ab initio scheme with level renormalization}},}\ }\href {\doibase 10.1103/PHYSREVB.99.245155/LA_3BAND_INTERACTION_CGW-SIC_DELTAMU.TXT} {\bibfield  {journal} {\bibinfo  {journal} {Phys. Rev. B}\ }\textbf {\bibinfo {volume} {99}},\ \bibinfo {pages} {245155} (\bibinfo {year} {2019})},\ \Eprint {http://arxiv.org/abs/1901.00763} {arXiv:1901.00763} \BibitemShut {NoStop}%
\bibitem [{\citenamefont {Hirayama}\ \emph {et~al.}(2018)\citenamefont {Hirayama}, \citenamefont {Yamaji}, \citenamefont {Misawa},\ and\ \citenamefont {Imada}}]{Hirayama2018}%
  \BibitemOpen
  \bibfield  {author} {\bibinfo {author} {\bibfnamefont {Motoaki}\ \bibnamefont {Hirayama}}, \bibinfo {author} {\bibfnamefont {Youhei}\ \bibnamefont {Yamaji}}, \bibinfo {author} {\bibfnamefont {Takahiro}\ \bibnamefont {Misawa}}, \ and\ \bibinfo {author} {\bibfnamefont {Masatoshi}\ \bibnamefont {Imada}},\ }\bibfield  {title} {\enquote {\bibinfo {title} {{Ab initio effective Hamiltonians for cuprate superconductors}},}\ }\href {\doibase 10.1103/PHYSREVB.98.134501/HIRAYAMA_CUPRATE_170820SUPPL.PDF} {\bibfield  {journal} {\bibinfo  {journal} {Phys. Rev. B}\ }\textbf {\bibinfo {volume} {98}},\ \bibinfo {pages} {134501} (\bibinfo {year} {2018})},\ \Eprint {http://arxiv.org/abs/1708.07498} {arXiv:1708.07498} \BibitemShut {NoStop}%
\bibitem [{\citenamefont {Weber}\ \emph {et~al.}(2010)\citenamefont {Weber}, \citenamefont {Haule},\ and\ \citenamefont {Kotliar}}]{Weber2010}%
  \BibitemOpen
  \bibfield  {author} {\bibinfo {author} {\bibfnamefont {C{\'{e}}dric}\ \bibnamefont {Weber}}, \bibinfo {author} {\bibfnamefont {Kristjan}\ \bibnamefont {Haule}}, \ and\ \bibinfo {author} {\bibfnamefont {Gabriel}\ \bibnamefont {Kotliar}},\ }\bibfield  {title} {\enquote {\bibinfo {title} {{Apical oxygens and correlation strength in electron- and hole-doped copper oxides}},}\ }\href {\doibase 10.1103/PHYSREVB.82.125107/FIGURES/20/THUMBNAIL} {\bibfield  {journal} {\bibinfo  {journal} {Phys. Rev. B - Condens. Matter Mater. Phys.}\ }\textbf {\bibinfo {volume} {82}},\ \bibinfo {pages} {125107} (\bibinfo {year} {2010})},\ \Eprint {http://arxiv.org/abs/1005.3100} {arXiv:1005.3100} \BibitemShut {NoStop}%
\bibitem [{\citenamefont {Werner}\ \emph {et~al.}(2015)\citenamefont {Werner}, \citenamefont {Sakuma}, \citenamefont {Nilsson},\ and\ \citenamefont {Aryasetiawan}}]{Werner2015}%
  \BibitemOpen
  \bibfield  {author} {\bibinfo {author} {\bibfnamefont {Philipp}\ \bibnamefont {Werner}}, \bibinfo {author} {\bibfnamefont {Rei}\ \bibnamefont {Sakuma}}, \bibinfo {author} {\bibfnamefont {Fredrik}\ \bibnamefont {Nilsson}}, \ and\ \bibinfo {author} {\bibfnamefont {Ferdi}\ \bibnamefont {Aryasetiawan}},\ }\bibfield  {title} {\enquote {\bibinfo {title} {{Dynamical screening in La2CuO4}},}\ }\href {\doibase 10.1103/PHYSREVB.91.125142/FIGURES/8/MEDIUM} {\bibfield  {journal} {\bibinfo  {journal} {Phys. Rev. B - Condens. Matter Mater. Phys.}\ }\textbf {\bibinfo {volume} {91}},\ \bibinfo {pages} {125142} (\bibinfo {year} {2015})},\ \Eprint {http://arxiv.org/abs/1411.3952} {arXiv:1411.3952} \BibitemShut {NoStop}%
\bibitem [{\citenamefont {Uchida}\ \emph {et~al.}(2017)\citenamefont {Uchida}, \citenamefont {{Kawasaki -}}, \citenamefont {Kim}, \citenamefont {{Kin-Ho Lee}}, \citenamefont {{Baek Kim -}}, \citenamefont {{Dundi Sri Chandana}}, \citenamefont {{Kumar Tiwari}}, \citenamefont {Saranya}, \citenamefont {Martins}, \citenamefont {Aichhorn},\ and\ \citenamefont {Biermann}}]{Uchida2017}%
  \BibitemOpen
  \bibfield  {author} {\bibinfo {author} {\bibfnamefont {Masaki}\ \bibnamefont {Uchida}}, \bibinfo {author} {\bibfnamefont {Masashi}\ \bibnamefont {{Kawasaki -}}}, \bibinfo {author} {\bibfnamefont {Heung-Sik}\ \bibnamefont {Kim}}, \bibinfo {author} {\bibfnamefont {Eric}\ \bibnamefont {{Kin-Ho Lee}}}, \bibinfo {author} {\bibfnamefont {Yong}\ \bibnamefont {{Baek Kim -}}}, \bibinfo {author} {\bibfnamefont {B}~\bibnamefont {{Dundi Sri Chandana}}}, \bibinfo {author} {\bibfnamefont {Jeetendra}\ \bibnamefont {{Kumar Tiwari}}}, \bibinfo {author} {\bibfnamefont {K}~\bibnamefont {Saranya}}, \bibinfo {author} {\bibfnamefont {C}~\bibnamefont {Martins}}, \bibinfo {author} {\bibfnamefont {M}~\bibnamefont {Aichhorn}}, \ and\ \bibinfo {author} {\bibfnamefont {S}~\bibnamefont {Biermann}},\ }\bibfield  {title} {\enquote {\bibinfo {title} {{Coulomb correlations in 4d and 5d oxides from first principles—or how spin–orbit materials choose their effective orbital degeneracies}},}\ }\href {\doibase 10.1088/1361-648X/AA648F} {\bibfield  {journal} {\bibinfo  {journal} {J. Phys. Condens. Matter}\ }\textbf {\bibinfo {volume} {29}},\ \bibinfo {pages} {263001} (\bibinfo {year} {2017})},\ \Eprint {http://arxiv.org/abs/1612.07546} {arXiv:1612.07546} \BibitemShut {NoStop}%
\bibitem [{\citenamefont {Li}\ \emph {et~al.}(2019)\citenamefont {Li}, \citenamefont {Lee}, \citenamefont {Wang}, \citenamefont {Osada}, \citenamefont {Crossley}, \citenamefont {Lee}, \citenamefont {Cui}, \citenamefont {Hikita},\ and\ \citenamefont {Hwang}}]{Li2019}%
  \BibitemOpen
  \bibfield  {author} {\bibinfo {author} {\bibfnamefont {Danfeng}\ \bibnamefont {Li}}, \bibinfo {author} {\bibfnamefont {Kyuho}\ \bibnamefont {Lee}}, \bibinfo {author} {\bibfnamefont {Bai~Yang}\ \bibnamefont {Wang}}, \bibinfo {author} {\bibfnamefont {Motoki}\ \bibnamefont {Osada}}, \bibinfo {author} {\bibfnamefont {Samuel}\ \bibnamefont {Crossley}}, \bibinfo {author} {\bibfnamefont {Hye~Ryoung}\ \bibnamefont {Lee}}, \bibinfo {author} {\bibfnamefont {Yi}~\bibnamefont {Cui}}, \bibinfo {author} {\bibfnamefont {Yasuyuki}\ \bibnamefont {Hikita}}, \ and\ \bibinfo {author} {\bibfnamefont {Harold~Y.}\ \bibnamefont {Hwang}},\ }\bibfield  {title} {\enquote {\bibinfo {title} {{Superconductivity in an infinite-layer nickelate}},}\ }\href {\doibase 10.1038/s41586-019-1496-5} {\bibfield  {journal} {\bibinfo  {journal} {Nature}\ }\textbf {\bibinfo {volume} {572}},\ \bibinfo {pages} {624--627} (\bibinfo {year} {2019})}\BibitemShut {NoStop}%
\bibitem [{\citenamefont {Sun}\ \emph {et~al.}(2023)\citenamefont {Sun}, \citenamefont {Huo}, \citenamefont {Hu}, \citenamefont {Li}, \citenamefont {Liu}, \citenamefont {Han}, \citenamefont {Tang}, \citenamefont {Mao}, \citenamefont {Yang}, \citenamefont {Wang}, \citenamefont {Cheng}, \citenamefont {Yao}, \citenamefont {Zhang},\ and\ \citenamefont {Wang}}]{Sun2023}%
  \BibitemOpen
  \bibfield  {author} {\bibinfo {author} {\bibfnamefont {Hualei}\ \bibnamefont {Sun}}, \bibinfo {author} {\bibfnamefont {Mengwu}\ \bibnamefont {Huo}}, \bibinfo {author} {\bibfnamefont {Xunwu}\ \bibnamefont {Hu}}, \bibinfo {author} {\bibfnamefont {Jingyuan}\ \bibnamefont {Li}}, \bibinfo {author} {\bibfnamefont {Zengjia}\ \bibnamefont {Liu}}, \bibinfo {author} {\bibfnamefont {Yifeng}\ \bibnamefont {Han}}, \bibinfo {author} {\bibfnamefont {Lingyun}\ \bibnamefont {Tang}}, \bibinfo {author} {\bibfnamefont {Zhongquan}\ \bibnamefont {Mao}}, \bibinfo {author} {\bibfnamefont {Pengtao}\ \bibnamefont {Yang}}, \bibinfo {author} {\bibfnamefont {Bosen}\ \bibnamefont {Wang}}, \bibinfo {author} {\bibfnamefont {Jinguang}\ \bibnamefont {Cheng}}, \bibinfo {author} {\bibfnamefont {Dao~Xin}\ \bibnamefont {Yao}}, \bibinfo {author} {\bibfnamefont {Guang~Ming}\ \bibnamefont {Zhang}}, \ and\ \bibinfo {author} {\bibfnamefont {Meng}\ \bibnamefont {Wang}},\ }\bibfield  {title} {\enquote {\bibinfo {title} {{Signatures of superconductivity near 80 K in a nickelate under high pressure}},}\ }\href {\doibase 10.1038/s41586-023-06408-7} {\bibfield  {journal} {\bibinfo  {journal} {Nature}\ }\textbf {\bibinfo {volume} {621}},\ \bibinfo {pages} {493--498} (\bibinfo {year} {2023})}\BibitemShut {NoStop}%
\bibitem [{\citenamefont {Saunderson}\ \emph {et~al.}(2020{\natexlab{a}})\citenamefont {Saunderson}, \citenamefont {Annett}, \citenamefont {{\'{U}}jfalussy}, \citenamefont {Csire},\ and\ \citenamefont {Gradhand}}]{Saunderson2020}%
  \BibitemOpen
  \bibfield  {author} {\bibinfo {author} {\bibfnamefont {Tom~G.}\ \bibnamefont {Saunderson}}, \bibinfo {author} {\bibfnamefont {James~F.}\ \bibnamefont {Annett}}, \bibinfo {author} {\bibfnamefont {Bal{\'{a}}zs}\ \bibnamefont {{\'{U}}jfalussy}}, \bibinfo {author} {\bibfnamefont {G{\'{a}}bor}\ \bibnamefont {Csire}}, \ and\ \bibinfo {author} {\bibfnamefont {Martin}\ \bibnamefont {Gradhand}},\ }\bibfield  {title} {\enquote {\bibinfo {title} {{Gap Anisotropy in Multiband Superconductors Based on Multiple Scattering Theory}},}\ }\href@noop {} {\bibfield  {journal} {\bibinfo  {journal} {Phys. Rev. B}\ }\textbf {\bibinfo {volume} {101}},\ \bibinfo {pages} {064510} (\bibinfo {year} {2020}{\natexlab{a}})}\BibitemShut {NoStop}%
\bibitem [{\citenamefont {Saunderson}\ \emph {et~al.}(2020{\natexlab{b}})\citenamefont {Saunderson}, \citenamefont {Győrgyp{\'{a}}l}, \citenamefont {Annett}, \citenamefont {Csire}, \citenamefont {{\'{U}}jfalussy},\ and\ \citenamefont {Gradhand}}]{Saunderson2020b}%
  \BibitemOpen
  \bibfield  {author} {\bibinfo {author} {\bibfnamefont {Tom~G.}\ \bibnamefont {Saunderson}}, \bibinfo {author} {\bibfnamefont {Zsolt}\ \bibnamefont {Győrgyp{\'{a}}l}}, \bibinfo {author} {\bibfnamefont {James~F.}\ \bibnamefont {Annett}}, \bibinfo {author} {\bibfnamefont {G{\'{a}}bor}\ \bibnamefont {Csire}}, \bibinfo {author} {\bibfnamefont {Bal{\'{a}}zs}\ \bibnamefont {{\'{U}}jfalussy}}, \ and\ \bibinfo {author} {\bibfnamefont {Martin}\ \bibnamefont {Gradhand}},\ }\bibfield  {title} {\enquote {\bibinfo {title} {{Real-space multiple scattering theory for superconductors with impurities}},}\ }\href {\doibase 10.1103/PhysRevB.102.245106} {\bibfield  {journal} {\bibinfo  {journal} {Phys. Rev. B}\ }\textbf {\bibinfo {volume} {102}},\ \bibinfo {pages} {245106} (\bibinfo {year} {2020}{\natexlab{b}})},\ \Eprint {http://arxiv.org/abs/2009.08766} {arXiv:2009.08766} \BibitemShut {NoStop}%
\bibitem [{\citenamefont {Saunderson}\ \emph {et~al.}(2022)\citenamefont {Saunderson}, \citenamefont {Annett}, \citenamefont {Csire},\ and\ \citenamefont {Gradhand}}]{Saunderson2022}%
  \BibitemOpen
  \bibfield  {author} {\bibinfo {author} {\bibfnamefont {Tom~G.}\ \bibnamefont {Saunderson}}, \bibinfo {author} {\bibfnamefont {James~F.}\ \bibnamefont {Annett}}, \bibinfo {author} {\bibfnamefont {G{\'{a}}bor}\ \bibnamefont {Csire}}, \ and\ \bibinfo {author} {\bibfnamefont {Martin}\ \bibnamefont {Gradhand}},\ }\bibfield  {title} {\enquote {\bibinfo {title} {{Full orbital decomposition of Yu-Shiba-Rusinov states based on first principles}},}\ }\href {\doibase 10.1103/PhysRevB.105.014424} {\bibfield  {journal} {\bibinfo  {journal} {Phys. Rev. B}\ }\textbf {\bibinfo {volume} {105}},\ \bibinfo {pages} {014424} (\bibinfo {year} {2022})}\BibitemShut {NoStop}%
\bibitem [{\citenamefont {Wu}\ \emph {et~al.}(2023)\citenamefont {Wu}, \citenamefont {Thill}, \citenamefont {Crosbie}, \citenamefont {Saunderson},\ and\ \citenamefont {Gradhand}}]{Wu2023}%
  \BibitemOpen
  \bibfield  {author} {\bibinfo {author} {\bibfnamefont {Ming-Hung}\ \bibnamefont {Wu}}, \bibinfo {author} {\bibfnamefont {Emma}\ \bibnamefont {Thill}}, \bibinfo {author} {\bibfnamefont {Jacob}\ \bibnamefont {Crosbie}}, \bibinfo {author} {\bibfnamefont {Tom~G.}\ \bibnamefont {Saunderson}}, \ and\ \bibinfo {author} {\bibfnamefont {Martin}\ \bibnamefont {Gradhand}},\ }\bibfield  {title} {\enquote {\bibinfo {title} {{Magnetic impurities on superconducting Pb surfaces}},}\ }\href {\doibase 10.1103/PhysRevB.107.094409} {\bibfield  {journal} {\bibinfo  {journal} {Phys. Rev. B}\ }\textbf {\bibinfo {volume} {107}},\ \bibinfo {pages} {094409} (\bibinfo {year} {2023})}\BibitemShut {NoStop}%
\bibitem [{\citenamefont {Csire}\ \emph {et~al.}(2018{\natexlab{a}})\citenamefont {Csire}, \citenamefont {De{\'{a}}k}, \citenamefont {Ny{\'{a}}ri}, \citenamefont {Ebert}, \citenamefont {Annett},\ and\ \citenamefont {{\'{U}}jfalussy}}]{Csire2018}%
  \BibitemOpen
  \bibfield  {author} {\bibinfo {author} {\bibfnamefont {G.}~\bibnamefont {Csire}}, \bibinfo {author} {\bibfnamefont {A.}~\bibnamefont {De{\'{a}}k}}, \bibinfo {author} {\bibfnamefont {B.}~\bibnamefont {Ny{\'{a}}ri}}, \bibinfo {author} {\bibfnamefont {H.}~\bibnamefont {Ebert}}, \bibinfo {author} {\bibfnamefont {J.~F.}\ \bibnamefont {Annett}}, \ and\ \bibinfo {author} {\bibfnamefont {B.}~\bibnamefont {{\'{U}}jfalussy}},\ }\bibfield  {title} {\enquote {\bibinfo {title} {{Relativistic spin-polarized KKR theory for superconducting heterostructures: Oscillating order parameter in the Au layer of Nb/Au/Fe trilayers}},}\ }\href {\doibase 10.1103/PhysRevB.97.024514} {\bibfield  {journal} {\bibinfo  {journal} {Phys. Rev. B}\ }\textbf {\bibinfo {volume} {97}},\ \bibinfo {pages} {024514} (\bibinfo {year} {2018}{\natexlab{a}})}\BibitemShut {NoStop}%
\bibitem [{\citenamefont {Csire}\ \emph {et~al.}(2018{\natexlab{b}})\citenamefont {Csire}, \citenamefont {{\'{U}}jfalussy},\ and\ \citenamefont {Annett}}]{Csire2018b}%
  \BibitemOpen
  \bibfield  {author} {\bibinfo {author} {\bibfnamefont {G{\'{a}}bor}\ \bibnamefont {Csire}}, \bibinfo {author} {\bibfnamefont {Bal{\'{a}}zs}\ \bibnamefont {{\'{U}}jfalussy}}, \ and\ \bibinfo {author} {\bibfnamefont {James~F.}\ \bibnamefont {Annett}},\ }\bibfield  {title} {\enquote {\bibinfo {title} {{Nonunitary triplet pairing in the noncentrosymmetric superconductor LaNiC2}},}\ }\href {\doibase 10.1140/EPJB/E2018-90095-7} {\bibfield  {journal} {\bibinfo  {journal} {Eur. Phys. J. B}\ }\textbf {\bibinfo {volume} {91}} (\bibinfo {year} {2018}{\natexlab{b}}),\ 10.1140/EPJB/E2018-90095-7}\BibitemShut {NoStop}%
\bibitem [{\citenamefont {Csire}\ \emph {et~al.}(2022)\citenamefont {Csire}, \citenamefont {Annett}, \citenamefont {Quintanilla},\ and\ \citenamefont {{\'{U}}jfalussy}}]{Csire2020a}%
  \BibitemOpen
  \bibfield  {author} {\bibinfo {author} {\bibfnamefont {G{\'{a}}bor}\ \bibnamefont {Csire}}, \bibinfo {author} {\bibfnamefont {James~F}\ \bibnamefont {Annett}}, \bibinfo {author} {\bibfnamefont {Jorge}\ \bibnamefont {Quintanilla}}, \ and\ \bibinfo {author} {\bibfnamefont {Bal{\'{a}}zs}\ \bibnamefont {{\'{U}}jfalussy}},\ }\bibfield  {title} {\enquote {\bibinfo {title} {{Magnetically-textured superconductivity in elemental Rhenium}},}\ }\href {\doibase 10.48550/arxiv.2005.05702} {\bibfield  {journal} {\bibinfo  {journal} {Phys. Rev. B}\ }\textbf {\bibinfo {volume} {106}},\ \bibinfo {pages} {L020501} (\bibinfo {year} {2022})},\ \Eprint {http://arxiv.org/abs/2005.05702} {arXiv:2005.05702} \BibitemShut {NoStop}%
\bibitem [{\citenamefont {Ghosh}\ \emph {et~al.}(2020)\citenamefont {Ghosh}, \citenamefont {Csire}, \citenamefont {Whittlesea}, \citenamefont {Annett}, \citenamefont {Gradhand}, \citenamefont {{\'{U}}jfalussy},\ and\ \citenamefont {Quintanilla}}]{Ghosh2020b}%
  \BibitemOpen
  \bibfield  {author} {\bibinfo {author} {\bibfnamefont {Sudeep~Kumar}\ \bibnamefont {Ghosh}}, \bibinfo {author} {\bibfnamefont {G{\'{a}}bor}\ \bibnamefont {Csire}}, \bibinfo {author} {\bibfnamefont {Philip}\ \bibnamefont {Whittlesea}}, \bibinfo {author} {\bibfnamefont {James~F.}\ \bibnamefont {Annett}}, \bibinfo {author} {\bibfnamefont {Martin}\ \bibnamefont {Gradhand}}, \bibinfo {author} {\bibfnamefont {Bal{\'{a}}zs}\ \bibnamefont {{\'{U}}jfalussy}}, \ and\ \bibinfo {author} {\bibfnamefont {Jorge}\ \bibnamefont {Quintanilla}},\ }\bibfield  {title} {\enquote {\bibinfo {title} {{Quantitative theory of triplet pairing in the unconventional superconductor LaNiGa2}},}\ }\href {\doibase 10.1103/PHYSREVB.101.100506/FIGURES/3/MEDIUM} {\bibfield  {journal} {\bibinfo  {journal} {Phys. Rev. B}\ }\textbf {\bibinfo {volume} {101}},\ \bibinfo {pages} {100506} (\bibinfo {year} {2020})},\ \Eprint {http://arxiv.org/abs/1912.08160} {arXiv:1912.08160} \BibitemShut {NoStop}%
\bibitem [{\citenamefont {R{\"{u}}{\ss}mann}\ and\ \citenamefont {Bl{\"{u}}gel}(2022{\natexlab{a}})}]{Rußmann2022b}%
  \BibitemOpen
  \bibfield  {author} {\bibinfo {author} {\bibfnamefont {Philipp}\ \bibnamefont {R{\"{u}}{\ss}mann}}\ and\ \bibinfo {author} {\bibfnamefont {Stefan}\ \bibnamefont {Bl{\"{u}}gel}},\ }\bibfield  {title} {\enquote {\bibinfo {title} {{Density functional Bogoliubov-de Gennes analysis of superconducting Nb and Nb(110) surfaces}},}\ }\href {\doibase 10.1103/PHYSREVB.105.125143/FIGURES/6/MEDIUM} {\bibfield  {journal} {\bibinfo  {journal} {Phys. Rev. B}\ }\textbf {\bibinfo {volume} {105}},\ \bibinfo {pages} {125143} (\bibinfo {year} {2022}{\natexlab{a}})},\ \Eprint {http://arxiv.org/abs/2110.01713} {arXiv:2110.01713} \BibitemShut {NoStop}%
\bibitem [{\citenamefont {R{\"{u}}{\ss}mann}\ \emph {et~al.}(2023)\citenamefont {R{\"{u}}{\ss}mann}, \citenamefont {Bahari}, \citenamefont {Bl{\"{u}}gel},\ and\ \citenamefont {Trauzettel}}]{Rußmann2023}%
  \BibitemOpen
  \bibfield  {author} {\bibinfo {author} {\bibfnamefont {Philipp}\ \bibnamefont {R{\"{u}}{\ss}mann}}, \bibinfo {author} {\bibfnamefont {Masoud}\ \bibnamefont {Bahari}}, \bibinfo {author} {\bibfnamefont {Stefan}\ \bibnamefont {Bl{\"{u}}gel}}, \ and\ \bibinfo {author} {\bibfnamefont {Bj{\"{o}}rn}\ \bibnamefont {Trauzettel}},\ }\bibfield  {title} {\enquote {\bibinfo {title} {{Interorbital Cooper pairing at finite energies in Rashba surface states}},}\ }\href {\doibase 10.1103/PHYSREVRESEARCH.5.043181/FIGURES/9/MEDIUM} {\bibfield  {journal} {\bibinfo  {journal} {Phys. Rev. Res.}\ }\textbf {\bibinfo {volume} {5}},\ \bibinfo {pages} {043181} (\bibinfo {year} {2023})},\ \Eprint {http://arxiv.org/abs/2307.13990} {arXiv:2307.13990} \BibitemShut {NoStop}%
\bibitem [{\citenamefont {Yamazaki}\ \emph {et~al.}(2025)\citenamefont {Yamazaki}, \citenamefont {Csire}, \citenamefont {Kucska}, \citenamefont {Shannon}, \citenamefont {Gyorffy}, \citenamefont {Takagi},\ and\ \citenamefont {{\'{U}}jfalussy}}]{Yamazaki2025}%
  \BibitemOpen
  \bibfield  {author} {\bibinfo {author} {\bibfnamefont {Hiroki}\ \bibnamefont {Yamazaki}}, \bibinfo {author} {\bibfnamefont {G{\'{a}}bor}\ \bibnamefont {Csire}}, \bibinfo {author} {\bibfnamefont {N{\'{o}}ra}\ \bibnamefont {Kucska}}, \bibinfo {author} {\bibfnamefont {Nic}\ \bibnamefont {Shannon}}, \bibinfo {author} {\bibfnamefont {Bal{\'{a}}zs}\ \bibnamefont {Gyorffy}}, \bibinfo {author} {\bibfnamefont {Hidenori}\ \bibnamefont {Takagi}}, \ and\ \bibinfo {author} {\bibfnamefont {Bal{\'{a}}zs}\ \bibnamefont {{\'{U}}jfalussy}},\ }\bibfield  {title} {\enquote {\bibinfo {title} {{Quantum Size Effects on Andreev Transport in Nb/Au/Nb Josephson Junctions: A Combined Ab Initio and Experimental Study}},}\ }\href {\doibase 10.1103/PHYSREVLETT.134.196002/SUPPMAT.PDF} {\bibfield  {journal} {\bibinfo  {journal} {Phys. Rev. Lett.}\ }\textbf {\bibinfo {volume} {134}},\ \bibinfo {pages} {196002} (\bibinfo {year} {2025})},\ \Eprint {http://arxiv.org/abs/2404.09784} {arXiv:2404.09784} \BibitemShut {NoStop}%
\bibitem [{\citenamefont {Reho}\ \emph {et~al.}(2024{\natexlab{a}})\citenamefont {Reho}, \citenamefont {Wittemeier}, \citenamefont {Kole}, \citenamefont {Ordej{\'{o}}n},\ and\ \citenamefont {Zanolli}}]{Reho2024}%
  \BibitemOpen
  \bibfield  {author} {\bibinfo {author} {\bibfnamefont {R.}~\bibnamefont {Reho}}, \bibinfo {author} {\bibfnamefont {N.}~\bibnamefont {Wittemeier}}, \bibinfo {author} {\bibfnamefont {A.~H.}\ \bibnamefont {Kole}}, \bibinfo {author} {\bibfnamefont {P.}~\bibnamefont {Ordej{\'{o}}n}}, \ and\ \bibinfo {author} {\bibfnamefont {Z.}~\bibnamefont {Zanolli}},\ }\bibfield  {title} {\enquote {\bibinfo {title} {{Density functional Bogoliubov-de Gennes theory for superconductors implemented in the SIESTA code}},}\ }\href {\doibase 10.1103/PHYSREVB.110.134505/FIGURES/15/MEDIUM} {\bibfield  {journal} {\bibinfo  {journal} {Phys. Rev. B}\ }\textbf {\bibinfo {volume} {110}},\ \bibinfo {pages} {134505} (\bibinfo {year} {2024}{\natexlab{a}})},\ \Eprint {http://arxiv.org/abs/2406.02022} {arXiv:2406.02022} \BibitemShut {NoStop}%
\bibitem [{\citenamefont {Ny{\'{a}}ri}\ \emph {et~al.}(2023)\citenamefont {Ny{\'{a}}ri}, \citenamefont {L{\'{a}}szl{\'{o}}ffy}, \citenamefont {Csire}, \citenamefont {Szunyogh},\ and\ \citenamefont {{\'{U}}jfalussy}}]{Nyari2023}%
  \BibitemOpen
  \bibfield  {author} {\bibinfo {author} {\bibfnamefont {Bendeg{\'{u}}z}\ \bibnamefont {Ny{\'{a}}ri}}, \bibinfo {author} {\bibfnamefont {Andr{\'{a}}s}\ \bibnamefont {L{\'{a}}szl{\'{o}}ffy}}, \bibinfo {author} {\bibfnamefont {G{\'{a}}bor}\ \bibnamefont {Csire}}, \bibinfo {author} {\bibfnamefont {L{\'{a}}szl{\'{o}}}\ \bibnamefont {Szunyogh}}, \ and\ \bibinfo {author} {\bibfnamefont {Bal{\'{a}}zs}\ \bibnamefont {{\'{U}}jfalussy}},\ }\bibfield  {title} {\enquote {\bibinfo {title} {{Topological superconductivity from first principles. I. Shiba band structure and topological edge states of artificial spin chains}},}\ }\href {\doibase 10.1103/PHYSREVB.108.134512/FIGURES/4/MEDIUM} {\bibfield  {journal} {\bibinfo  {journal} {Phys. Rev. B}\ }\textbf {\bibinfo {volume} {108}},\ \bibinfo {pages} {134512} (\bibinfo {year} {2023})},\ \Eprint {http://arxiv.org/abs/2308.13824} {arXiv:2308.13824} \BibitemShut {NoStop}%
\bibitem [{\citenamefont {L{\'{a}}szl{\'{o}}ffy}\ \emph {et~al.}(2023)\citenamefont {L{\'{a}}szl{\'{o}}ffy}, \citenamefont {Ny{\'{a}}ri}, \citenamefont {Csire}, \citenamefont {Szunyogh},\ and\ \citenamefont {{\'{U}}jfalussy}}]{Laszloffy2023}%
  \BibitemOpen
  \bibfield  {author} {\bibinfo {author} {\bibfnamefont {Andr{\'{a}}s}\ \bibnamefont {L{\'{a}}szl{\'{o}}ffy}}, \bibinfo {author} {\bibfnamefont {Bendeg{\'{u}}z}\ \bibnamefont {Ny{\'{a}}ri}}, \bibinfo {author} {\bibfnamefont {G{\'{a}}bor}\ \bibnamefont {Csire}}, \bibinfo {author} {\bibfnamefont {L{\'{a}}szl{\'{o}}}\ \bibnamefont {Szunyogh}}, \ and\ \bibinfo {author} {\bibfnamefont {Bal{\'{a}}zs}\ \bibnamefont {{\'{U}}jfalussy}},\ }\bibfield  {title} {\enquote {\bibinfo {title} {{Topological superconductivity from first principles. II. Effects from manipulation of spin spirals: Topological fragmentation, braiding, and quasi-Majorana bound states}},}\ }\href {\doibase 10.1103/PHYSREVB.108.134513/FIGURES/9/MEDIUM} {\bibfield  {journal} {\bibinfo  {journal} {Phys. Rev. B}\ }\textbf {\bibinfo {volume} {108}},\ \bibinfo {pages} {134513} (\bibinfo {year} {2023})}\BibitemShut {NoStop}%
\bibitem [{\citenamefont {R{\"{u}}{\ss}mann}\ and\ \citenamefont {Bl{\"{u}}gel}(2022{\natexlab{b}})}]{Rußmann2022a}%
  \BibitemOpen
  \bibfield  {author} {\bibinfo {author} {\bibfnamefont {Philipp}\ \bibnamefont {R{\"{u}}{\ss}mann}}\ and\ \bibinfo {author} {\bibfnamefont {Stefan}\ \bibnamefont {Bl{\"{u}}gel}},\ }\bibfield  {title} {\enquote {\bibinfo {title} {{Proximity induced superconductivity in a topological insulator}},}\ }\href {\doibase 10.48550/arxiv.2208.14289} {\  (\bibinfo {year} {2022}{\natexlab{b}}),\ 10.48550/arxiv.2208.14289},\ \Eprint {http://arxiv.org/abs/2208.14289} {arXiv:2208.14289} \BibitemShut {NoStop}%
\bibitem [{\citenamefont {Reho}\ \emph {et~al.}(2024{\natexlab{b}})\citenamefont {Reho}, \citenamefont {Botello-M{\'{e}}ndez},\ and\ \citenamefont {Zanolli}}]{Reho2024a}%
  \BibitemOpen
  \bibfield  {author} {\bibinfo {author} {\bibfnamefont {R.}~\bibnamefont {Reho}}, \bibinfo {author} {\bibfnamefont {A.~R.}\ \bibnamefont {Botello-M{\'{e}}ndez}}, \ and\ \bibinfo {author} {\bibfnamefont {Zeila}\ \bibnamefont {Zanolli}},\ }\bibfield  {title} {\enquote {\bibinfo {title} {{Ab initio study of Proximity-Induced Superconductivity in PbTe/Pb heterostructures}},}\ }\href {https://arxiv.org/abs/2412.01749v1} {\  (\bibinfo {year} {2024}{\natexlab{b}})},\ \Eprint {http://arxiv.org/abs/2412.01749} {arXiv:2412.01749} \BibitemShut {NoStop}%
\bibitem [{\citenamefont {Park}\ \emph {et~al.}(2020)\citenamefont {Park}, \citenamefont {Csire},\ and\ \citenamefont {Ujfalussy}}]{Park2020a}%
  \BibitemOpen
  \bibfield  {author} {\bibinfo {author} {\bibfnamefont {Kyungwha}\ \bibnamefont {Park}}, \bibinfo {author} {\bibfnamefont {Gabor}\ \bibnamefont {Csire}}, \ and\ \bibinfo {author} {\bibfnamefont {Balazs}\ \bibnamefont {Ujfalussy}},\ }\bibfield  {title} {\enquote {\bibinfo {title} {{Proximity effect in a superconductor-topological insulator heterostructure based on first principles}},}\ }\href {\doibase 10.1103/PHYSREVB.102.134504/FIGURES/5/MEDIUM} {\bibfield  {journal} {\bibinfo  {journal} {Phys. Rev. B}\ }\textbf {\bibinfo {volume} {102}},\ \bibinfo {pages} {134504} (\bibinfo {year} {2020})},\ \Eprint {http://arxiv.org/abs/2005.02570} {arXiv:2005.02570} \BibitemShut {NoStop}%
\bibitem [{Sup()}]{Supplementary}%
  \BibitemOpen
  \href@noop {} {}\bibinfo {note} {See Supplemental Material}\BibitemShut {NoStop}%
\bibitem [{\citenamefont {Nelmes}\ \emph {et~al.}(1990)\citenamefont {Nelmes}, \citenamefont {Wilding}, \citenamefont {Hatton}, \citenamefont {Caignaert}, \citenamefont {Raveau}, \citenamefont {McMahon},\ and\ \citenamefont {Piltz}}]{Nelmes1990}%
  \BibitemOpen
  \bibfield  {author} {\bibinfo {author} {\bibfnamefont {R.~J.}\ \bibnamefont {Nelmes}}, \bibinfo {author} {\bibfnamefont {N.~B.}\ \bibnamefont {Wilding}}, \bibinfo {author} {\bibfnamefont {P.~D.}\ \bibnamefont {Hatton}}, \bibinfo {author} {\bibfnamefont {V.}~\bibnamefont {Caignaert}}, \bibinfo {author} {\bibfnamefont {B.}~\bibnamefont {Raveau}}, \bibinfo {author} {\bibfnamefont {M.~I.}\ \bibnamefont {McMahon}}, \ and\ \bibinfo {author} {\bibfnamefont {R.~O.}\ \bibnamefont {Piltz}},\ }\bibfield  {title} {\enquote {\bibinfo {title} {{Pressure dependence of the structure of La-Sr-Cu-O}},}\ }\href {\doibase 10.1016/0921-4534(90)90413-9} {\bibfield  {journal} {\bibinfo  {journal} {Phys. C Supercond.}\ }\textbf {\bibinfo {volume} {166}},\ \bibinfo {pages} {329--333} (\bibinfo {year} {1990})}\BibitemShut {NoStop}%
\bibitem [{\citenamefont {Hayward}\ and\ \citenamefont {Rosseinsky}(2003)}]{Hayward2003}%
  \BibitemOpen
  \bibfield  {author} {\bibinfo {author} {\bibfnamefont {M.~A.}\ \bibnamefont {Hayward}}\ and\ \bibinfo {author} {\bibfnamefont {M.~J.}\ \bibnamefont {Rosseinsky}},\ }\bibfield  {title} {\enquote {\bibinfo {title} {{Synthesis of the infinite layer Ni(I) phase NdNiO2+x by low temperature reduction of NdNiO3 with sodium hydride}},}\ }\href {\doibase 10.1016/S1293-2558(03)00111-0} {\bibfield  {journal} {\bibinfo  {journal} {Solid State Sci.}\ }\textbf {\bibinfo {volume} {5}},\ \bibinfo {pages} {839--850} (\bibinfo {year} {2003})}\BibitemShut {NoStop}%
\bibitem [{\citenamefont {Radaelli}\ \emph {et~al.}(1994)\citenamefont {Radaelli}, \citenamefont {Hinks}, \citenamefont {Mitchell}, \citenamefont {Hunter}, \citenamefont {Wagner}, \citenamefont {Dabrowski}, \citenamefont {Vandervoort}, \citenamefont {Viswanathan},\ and\ \citenamefont {Jorgensen}}]{Radaelli1994}%
  \BibitemOpen
  \bibfield  {author} {\bibinfo {author} {\bibfnamefont {P.~G.}\ \bibnamefont {Radaelli}}, \bibinfo {author} {\bibfnamefont {D.~G.}\ \bibnamefont {Hinks}}, \bibinfo {author} {\bibfnamefont {A.~W.}\ \bibnamefont {Mitchell}}, \bibinfo {author} {\bibfnamefont {B.~A.}\ \bibnamefont {Hunter}}, \bibinfo {author} {\bibfnamefont {J.~L.}\ \bibnamefont {Wagner}}, \bibinfo {author} {\bibfnamefont {B.}~\bibnamefont {Dabrowski}}, \bibinfo {author} {\bibfnamefont {K.~G.}\ \bibnamefont {Vandervoort}}, \bibinfo {author} {\bibfnamefont {H.~K.}\ \bibnamefont {Viswanathan}}, \ and\ \bibinfo {author} {\bibfnamefont {J.~D.}\ \bibnamefont {Jorgensen}},\ }\bibfield  {title} {\enquote {\bibinfo {title} {{Structural and superconducting properties of La2SrCuO4 as a function of Sr content}},}\ }\href {\doibase 10.1103/PhysRevB.49.4163} {\bibfield  {journal} {\bibinfo  {journal} {Phys. Rev. B}\ }\textbf {\bibinfo {volume} {49}},\ \bibinfo {pages} {4163} (\bibinfo {year} {1994})}\BibitemShut {NoStop}%
\bibitem [{\citenamefont {Naito}\ \emph {et~al.}(2018)\citenamefont {Naito}, \citenamefont {Sato}, \citenamefont {Tsukada},\ and\ \citenamefont {Yamamoto}}]{Naito2018}%
  \BibitemOpen
  \bibfield  {author} {\bibinfo {author} {\bibfnamefont {Michio}\ \bibnamefont {Naito}}, \bibinfo {author} {\bibfnamefont {Hisashi}\ \bibnamefont {Sato}}, \bibinfo {author} {\bibfnamefont {Akio}\ \bibnamefont {Tsukada}}, \ and\ \bibinfo {author} {\bibfnamefont {Hideki}\ \bibnamefont {Yamamoto}},\ }\bibfield  {title} {\enquote {\bibinfo {title} {{Epitaxial effects in thin films of high-Tc cuprates with the K2NiF4 structure}},}\ }\href {\doibase 10.1016/J.PHYSC.2017.11.010} {\bibfield  {journal} {\bibinfo  {journal} {Phys. C Supercond. its Appl.}\ }\textbf {\bibinfo {volume} {546}},\ \bibinfo {pages} {84--114} (\bibinfo {year} {2018})}\BibitemShut {NoStop}%
\end{thebibliography}%

%@misc{Supplementary,
%Note={See Supplemental Material, which also contains %Refs.~\cite{Csire2018,Capelle1999,Capelle1999a,Zabloudil2004,Gradhand2009,Ebert1996,%Ebert2016,Saunderson2020,Saunderson2020b,Saunderson2022,Vosko1980,Chirolli2021,Zabloudil2004}}
%}

\end{document}